\documentclass[aps,prd,preprint]{revtex4}
\usepackage{amsfonts}
\usepackage{amssymb}
\usepackage{amsmath}
\newcommand{\be}{\begin{equation}}
\newcommand{\ee}{\end{equation}}
\newcommand{\bea}{\begin{eqnarray}}
\newcommand{\eea}{\end{eqnarray}}
\newcommand{\bwt}{\begin{widetext}}
\newcommand{\ewt}{\end{widetext}}
\begin{document}
\title{Holographic Currents and Chern-Simons Terms}
\author{T.E. Clark}
\email[e-mail address:]{clarkt@purdue.edu}
\affiliation{Department of Physics,\\
 Purdue University,\\
 West Lafayette, IN 47907-2036, U.S.A.}
\author{S.T. Love}
\email[e-mail address:]{loves@purdue.edu}
\affiliation{Department of Physics,\\
 Purdue University,\\
 West Lafayette, IN 47907-2036, U.S.A.}
\author{T. ter Veldhuis}
\email[e-mail address:]{terveldhuis@macalester.edu}
\affiliation{Department of Physics \& Astronomy,\\
 Macalester College,\\
 Saint Paul, MN 55105-1899, U.S.A.}

\vspace*{1.0in.}

\begin{abstract}
Holographic currents and their associated Ward identities are derived in the framework of gravity/gauge duality.  Holographic improvements of the energy-momentum tensor and $R$-symmetry current which are consistent with the Ward identities are displayed.  The effects of specific string loop corrections to the bulk action are included as four derivative effective lagrangian terms and their contributions to the trace and $R$-symmetry anomalies of the boundary theory are determined.  As an example, the construction is applied to the ${\cal N}=2$ conformal supergravity which is taken to be dual to a boundary $SU(N)\times SU(N)$, ${\cal N}=1$ superconformal field theory.
\end{abstract}

\maketitle

\section{Introduction \label{intro}}

Type IIB string theory on an $AdS_5 \times Y_5$ background, where $Y_5$ is a Sasaki-Einstein manifold, for large radii and with a stack of $N$-$D3$-branes at the tip of the conical singularity is approximated by a ${\cal N}=2$ conformal supergravity which in turn is dual to a ${\cal N}=1$ superconformal field theory on the Minkowski space-time $M_4$ boundary of the $AdS_5$ space \cite{Maldacena:1997re}-\cite{Gubser:1998bc}.  The values of the central charges of the superconformal field theory (SCFT), $a$ and $c$, designate different boundary field theories dual to the bulk $AdS$ supergravity \cite{Anselmi:1997am}.  
Indeed the energy-momentum tensor trace and the $R$-symmetry current anomalies calculated holographically in the supergravity theory and obtained in terms of the central charges in the boundary field theory have been shown to match in leading order \cite{Henningson:1998gx}, while imposing subleading order matching provides a measure of the string loop corrections to the supergravity action through the value of the coefficient governing the higher derivative action terms \cite{Nojiri:1999mh}-\cite{Cremonini:2008tw}.

The trace anomaly of the boundary ${\cal N}=1$ SCFT in the presence of background gravitational and $U(1)_R$ gauge fields  and in the absence of sources related to any other global symmetries has the general form \cite{Anselmi:1997am}
\be
\theta^\mu_{~\mu} = \frac{1}{16\pi^2} \left[ c W^2 -a {R}_{\rm GB}^2 \right] + \frac{c}{6\pi^2} F_{\mu\nu} F^{\mu\nu},
\ee
where $W_{\mu\nu\rho\sigma}= R_{\mu\nu\rho\sigma} +1/(d-2) [g_{\rho\mu} R_{\nu\sigma} -g_{\rho\nu} R_{\mu\sigma} -g_{\sigma\mu} R_{\nu\rho} +g_{\sigma\nu} R_{\mu\rho} ]-1/(d-2)(d-1) (g_{\rho\mu}g_{\nu\sigma}-g_{\rho\nu}g_{\mu\sigma}) R$ is the Weyl tensor in $d$-dimensional space-time with $R_{\mu\nu\rho\sigma}$ the Riemann tensor and ${R_{\rm GB}}^2$ the Einstein-Gauss-Bonnet invariant ${R_{\rm GB}}^2 =R_{\mu\nu\rho\sigma}^2 -4 R_{\mu\nu}^2 +R^2$ in all dimensions.  $F_{\mu\nu}$ is the $U(1)_R$ gauge field strength tensor.  Expanding the gravitational contributions to the anomaly yields
\be
\theta^\mu_{~\mu} = \frac{1}{16\pi^2} \left[ 2(2a -c) R_{\mu\nu} R^{\mu\nu} + \frac{1}{3}(c-3a) R^2 +(c-a) R_{\mu\nu\rho\sigma}  R^{\mu\nu\rho\sigma} \right] +\frac{c}{6\pi^2} F_{\mu\nu} F^{\mu\nu} .
\label{SCFTtraceanomaly}
\ee
For a $R$-current defined so that it lies in the same ${\cal N}=1$ SUSY supercurrent multiplet as the energy-momentum tensor and with the trace and $R$-current anomalies also appearing in their own SUSY anomaly multiplet, then the $R$-symmetry current anomaly is given in terms of the same $a$ and $c$ coefficients and has the form \cite{Anselmi:1997am}
\be
\partial_\mu (\sqrt{-g} R^\mu )= \frac{1}{24 \pi^2} (c-a) \epsilon^{\mu\nu\rho\sigma} R^{\xi\zeta}_{~~~\mu\nu} R_{\xi\zeta\rho\sigma} + \frac{1}{18 \pi^2} (5a-3c) \epsilon^{\mu\nu\rho\sigma} F_{\mu\nu} F_{\rho\sigma} .
\label{SCFTRanomaly}
\ee
In order for these currents, along with the SUSY current, to lie in the same ${\cal N}=1
$ supermultiplet, they must have appropriate normalization and improvements.  If the $R$-charges are defined differently, say by a rescaling by $r$, $R\rightarrow rR$, then the contribution to the $F\tilde{F}$ term will be scaled as $r^3$ while the $R\tilde{R}$ contribution to the anomaly triangle will be scaled only linearly with $r$.  Furthermore, if the overall normalization of the $R$-current is also scaled so that $R^\mu \rightarrow \lambda R^\mu$, then the two terms on the right hand side of the divergence equation (\ref{SCFTRanomaly})get modified by $\lambda r$ and $\lambda r^3$ factors respectively.  These factors will be necessary in matching the holographically defined boundary currents to the conventional SCFT supercurrent components.

The field theory determination of these anomaly coefficients is \cite{Anselmi:1997am}
\be
a= \frac{3}{32} \left( 3 {\rm Tr}{R^3} - {\rm Tr}{R} \right)  \qquad , \qquad c= \frac{1}{32} \left( 9 {\rm Tr}{R^3} - 5 {\rm Tr}{R} \right)  ,
\ee
where $R$ is the $R$-symmetry charge of the fields and the trace is over all chiral fermion fields.  Inverting these expressions yields the linear and cubic nature of their triangle contributions ${\rm Tr}[R] =-16 (c-a)$ and ${\rm Tr}[R^3] =16(5a-3c)/9$.  The leading $N^2$ contribution to the anomaly coefficients implies that $c=a$ for the class of models considered here.  Hence the gravitational field contribution to the trace anomaly occurs only through the Ricci tensor and scalar curvature terms and not the Riemann tensor term, while there is no leading order gravitational contribution to the $R$-symmetry anomaly.  The subleading large $N$ contributions to both anomalies are reflected not only in the $(c-a)\neq 0$ difference of the central charges but also in the subleading corrections to each individually. 

On the supergravity side, the leading order contributions to the anomalies are given by the 2 derivative terms in the bosonic part of the supersymmetric action
\be
\Gamma_{\rm Leading}= \int d^5 x \det{E} \left[ \frac{1}{2\kappa}R + \Lambda -\frac{Z_A}{4} F_{MN} F^{MN} + C E^{MNRST} F_{MN} F_{RS} A_T \right],
\ee
where $\det{E}\, E^{MNRST}= \epsilon^{MNRST}$ and $E_M^A$ is the gravitational f\"unfbein. 
The $AdS_5$ gravitational constant is $2\kappa = 16\pi G_5$ and the cosmological constant is $\Lambda = -6/\kappa R_5^2$ with $R_5$ the $AdS_5$ radius of curvature. The $U(1)_R$ field strength tensor $F_{MN} = \partial_M A_N -\partial_N A_M$ while the re-scaled gauge field wavefunction normalization factor $Z_A$ is determined by matching the holographic energy-momentum tensor anomaly to that of the boundary SCFT as given in equation (\ref{SCFTtraceanomaly}).  The ${\cal N}=2$ SUSY determines the Chern-Simons coefficient (see Appendix A)
\be
C= \frac{(Z_A 2\kappa )^{3/2}}{24 \kappa \sqrt{3}}.
\label{LeadingC}
\ee
The leading order trace anomaly was found in reference \cite{Henningson:1998gx} in the absence of the $U(1)_R$ field strength.  After its inclusion, the trace anomaly is
\be
\theta^\mu_{~~\mu} = \left(\frac{1}{8}\frac{R_5^3}{\kappa}\right)\left[  R_{\mu\nu}R^{\mu\nu}-\frac{1}{3}  R^2 \right] 
  +\left( \frac{R_5 Z_A}{4}\right) F_{\mu\nu} F^{\mu\nu} .
\label{holographictraceanomalyLeading}
\ee
Comparing this to equation (\ref{SCFTtraceanomaly}) yields the leading order results that $a=c$ and identifies the gravitational constant as $R_5^3 /\kappa =a/\pi^2$.  Requiring consistency for the $F^2$ contribution yields that the wavefunction normalization factor satisfies $Z_A R_5 =2a/3\pi^2$.  The direct variation of the action with respect to $R$-symmetry transformations, $\delta_R A_M = \partial_M \Omega$, yields the divergence of the holographic $R$-current in leading order as
\be
\delta_R \Gamma_{\rm Leading} =-C \int_{M_4} d^4 x\,\, \Omega (x) \epsilon^{\mu\nu\rho\sigma} F_{\mu\nu} F_{\rho\sigma} .
\ee
$AdS / CFT$ duality implies that the holographic $R$-current anomaly coefficient $C$ is
\be
\lambda r^3 C= \frac{1}{18 \pi^2} (5a -3c)=\frac{a}{9\pi^2}  .
\label{CmatchingLeading}
\ee
Inserting the definition of $C$ from equation (\ref{LeadingC}) in terms of $a$ and $c$ requires the normalization of the charges and current to obey $\lambda r^3 = 3$ in order for the $R$-current to belong to the ${\cal N}=1$ supercurrent multiplet.

The subleading contributions to the anomalies arise due to higher order 4 derivative terms in the action coming from string loop effects along with subleading corrections to the coefficients of the 2 derivative terms \cite{Nojiri:1999mh}-\cite{Cremonini:2008tw}.  The $R$-symmetry anomaly arises directly from the variation of the 5 dimensional mixed $U(1)_R$ gauge field and gravitational field Chern-Simons term as
\be
\Gamma_{RRA} = -D \int d^5 x \det{E}\, E^{MNRST} R^X_{~~YMN} R^Y_{~~XRS} A_T ,
\ee
where $D\equiv -\gamma R_5^3 /3$ and $\gamma$ sets the scale of the SUSY completion of the 4 derivative contributions to the action.  In the notation of reference \cite{Cremonini:2008tw} $16 \pi G_5 \gamma = c_2/ 192$.  The $U(1)_R$ transformation, $\delta_R A_T = \partial_T \Omega$, yields the gravitational contribution to the $R$-symmetry anomaly as
\be
\delta_R \Gamma_{RRA} =-D \int_{M_4} d^4 x\,\, \Omega (x) \epsilon^{\mu\nu\rho\sigma} R^{\xi\zeta}_{~~~\mu\nu} R_{\xi\zeta\rho\sigma} .
\ee
Exploiting $AdS / CFT$ duality, the boundary SCFT result for the gravitational contribution to the $R$-anomaly is such that the holographic $R$-current and $R$-charges must include the rescaling by $\lambda$ and $r$ in order for the $R$-current to fit in the supercurrent with the holographic energy-momentum tensor, so that
\be
\lambda r D=\frac{1}{24 \pi^2} (c-a) .
\label{Dmatching}
\ee
As in leading order case, the 2-derivative pure $U(1)_R$ gauge Chern-Simons term is responsible for the gauge field contribution to the $R$-anomaly.  With an $AdS$ action term of the form  \cite{Witten:1998qj}, \cite{Klebanov:2002gr}
\be
\Gamma_{FFA} = C \int d^5 x \det{E}\, E^{MNRST} F_{MN} F_{RS} A_T ,
\ee
where $C$ is fixed by supersymmetry \cite{Cremonini:2008tw}   through subleading order (see Appendix A) to be $C= 2 R_5^3 [1+ 12 \gamma (2\kappa)/R_5^2 ]/27 (2\kappa)$.  As previously, its contribution to the $R$-anomaly is 
\be
\delta_R \Gamma_{FFA} =-C \int_{M_4} d^4 x\,\, \Omega (x) \epsilon^{\mu\nu\rho\sigma} F_{\mu\nu} F_{\rho\sigma} .
\ee
$AdS / CFT$ duality implies that the $R$-anomaly anomaly coefficient $C$ is
\be
\lambda r^3 C= \frac{1}{18 \pi^2} (5a -3c)  .
\label{Cmatching}
\ee
Note that $(c-a)/a = 8D/(3 r^2 C+12D)=8 D/3 r^2 C$ to subleading order.

On the other hand, the SUSY completion of the mixed Chern-Simons term requires many more 4 derivative terms to be added to the supergravity action \cite{Nojiri:1999mh}-\cite{Cremonini:2008tw} albeit with relative coefficients determined by the supergravity transformations and overall coefficient fixed by $\gamma$ \cite{Cremonini:2008tw}.  In the SUSY case discussed here, the Weyl tensor squared term has been shown \cite{Cremonini:2008tw} to be the source of the corresponding subleading pure gravitational corrections to the trace anomaly.  The two independent gravitational tensor couplings to the $U(1)_R$ field strength bilinear along with the three field strength covariant derivative bilinears provide the subleading field strength squared corrections as shown in Appendix A for arbitrary coefficients.  (Indeed the coupling of the Weyl tensor to the field strength, $F_{MN}F_{RS} W^{MNRS}$, does not contribute to the anomaly while the remaining couplings to the Ricci tensor and scalar curvature do contribute.)  For the conformal supergravity values of the coefficients \cite{Cremonini:2008tw} this action reads
\bea
\Gamma_{\rm Subleading} &=& \gamma \int d^5 x \det{E} \biggl[W_{MNRS} W^{MNRS}+\frac{R_5}{3} E^{MNRST} R^X_{~~YMN} R^Y_{~~XRS} A_T    \cr
 & &\left. \qquad\qquad\qquad\qquad +\frac{2}{9}R_5^2 F_{MN}F^{MN}R -\frac{32}{9} R_5^2 F_{MR} F^{~~R}_{N} R^{MN} \right. \cr
 & &\left. -\frac{16}{3} R_5^2 F_{MN} \nabla^N \nabla_R F^{MR} -\frac{8}{3} R_5^2\nabla^M F^{NR} \nabla_M F_{NR}  -\frac{8}{3} R_5^2 \nabla^M F^{NR} \nabla_N F_{MR} \right. \cr 
 & &  \qquad\qquad\qquad\qquad\qquad\qquad + {\rm SUSY ~Completion}\biggr] .
\label{subleadingaction}
\eea
The remaining SUSY completion terms do not contribute to either anomaly.
The holographic trace anomaly in the general non-SUSY case is considered in Appendix A.  Specializing the results of Appendix A to the ${\cal N}=2$ conformal supergravity case, the holographic trace anomaly has the form
\bea
\theta^\mu_{~~\mu} &=&  \left(-\frac{1}{24}\frac{R_5^3}{\kappa} -\frac{1}{3}R_5 \gamma \right) R^2 +\left(\frac{1}{8}\frac{R_5^3}{\kappa} +2R_5 \gamma \right) R_{\mu\nu}R^{\mu\nu} -R_5 \gamma R_{\mu\nu\rho\sigma} R^{\mu\nu\rho\sigma} \cr
 & &\qquad\qquad\qquad +\left( \frac{R_5 Z_A}{4}-\frac{160}{9}R_5 \gamma\right) F_{\mu\nu} F^{\mu\nu} .
\label{holographictraceanomaly}
\eea

Using $AdS / CFT$ duality to match the trace anomaly in the SCFT form, equation (\ref{SCFTtraceanomaly}), to the holographic trace anomaly, equation (\ref{holographictraceanomaly}), leads to the values for the gravitational constant $\kappa$ and the scale of the subleading corrections $\gamma$
\bea
\frac{R_5^3}{\kappa} &=& \frac{a}{\pi^2} \cr
\gamma R_5 &=& -\frac{(c-a)}{16\pi^2} .
\eea
Consistency also requires the value of the coefficient of the Riemann tensor squared to be given as $(c-a)/16 \pi^2$, in agreement with the above results.  In addition, it is in agreement with the conformal supergravity symmetry constraints relating the subleading gravitational contribution to the trace anomaly to just the Weyl tensor squared as given in equation (\ref{subleadingaction}).  The $U(1)_R$ field strength contribution to the anomaly involves only the coefficient $c$.  This part of the trace anomaly fixes the wavefunction renormalization of the $U(1)_R$ gauge field to be
\be
Z_A R_5 = \frac{2}{3} \frac{a}{\pi^2} + \frac{10}{9} \frac{(c-a)}{\pi^2} .
\label{wavefunction}
\ee

The ${\cal N}=2$ supergravity constraints require the supersymmetric values of $C$ and $D$ to be given by
\bea
C&=& \frac{2 R_5^3}{27 (2\kappa)} \left[ 1 + 12 \frac{\gamma (2\kappa)}{R_5^2}\right] = \frac{1}{3}\frac{(5a -3c)}{18 \pi^2} \cr
D&=&-\frac{\gamma R_5}{3} = \frac{1}{2} \frac{(c-a)}{24 \pi^2} ,
\eea
where the second equality employs the relationship of the supergravity coefficients to $a$ and $c$ from the trace anomaly above.  Turning to the $R$-symmetry $AdS/CFT$ anomaly matching equations (\ref{Dmatching}) and (\ref{Cmatching}), the Chern-Simons coefficients imply the need to normalize the $R$-charges and current so that $r=\sqrt{3/2}$ and $\lambda = \sqrt{8/3}$ for the holographic $R$-current to belong to the supercurrent.

As an example, taking the above mentioned Sasaki-Einstein manifold to be $Y_5=T^{1,1}$, these anomaly matching results can be applied to the large $N$ strong t'Hooft coupling superconformal  $SU(N)\times SU(N)$ SUSY gauge theory with a pair of bi-fundamental $(N,\bar{N})$ chiral matter fields $A_1$ and $A_2$ and a pair of anti-bi-fundamental $(\bar{N}, N)$ chiral matter fields $B_1$ and $B_2$ \cite{Klebanov:1998hh}-\cite{Klebanov:2002gr}. The central charges calculated in the boundary field theory determine the two supergravity parameters related to the gravitational constant $\kappa$ and $\gamma$ which sets the scale for the four derivative terms.  The $A$ and $B$ chiral superfields have $R=1/2$, so the $\tilde{A}$ and $\tilde{B}$ matter fermions have $R=-1/2$.  The $2(N^2 -1)$ gluinos have $R=1$.  Hence the boundary SCFT has
\be
a= \frac{27}{64}N^2 -\frac{3}{8} \qquad , \qquad c= \frac{27}{64}N^2 -\frac{1}{4}  ,
\ee
with $(c-a) = 1/8$.  Thus the gravitational constant is seen to be
\be
\frac{R_5^3}{\kappa} = \frac{27}{64 \pi^2}N^2 \left(1-\frac{8}{9N^2}\right),
\ee
and the wave function renormalization $Z_A$ is found to be 
\be
Z_A R_5=\frac{9}{32\pi^2}N^2\left( 1-\frac{32}{81 N^2} \right) ,
\ee
while the coefficients of the 4 derivative terms are given by $\gamma$ and are found to be subleading
\be
\gamma R_5 = -\frac{1}{128 \pi^2} .
\ee

The anomaly matching results aside, the purpose of this paper is to construct the energy-momentum tensor and $R$-symmetry current in the case that the subleading mixed Chern-Simons term is added to the leading order supergravity action.  This truncated 4 derivative action will provide the leading gravitational contributions to each anomaly (leading $N^2$ contributions with $c=a$ in the trace anomaly case and subleading $N^0$ contributions with $c \neq a$ in the $R$-anomaly case).  The pure $U(1)_R$ gauge field contributions are leading order $N^2$ terms for each anomaly.  In section \ref{section2}, the near boundary field equations are solved and the abbrreviated action is holographically renormalized through the addition of boundary counter-terms \cite{de Haro:2000xn}-\cite{Skenderis:2002wp}.  In section \ref{section3} the Brown-York energy-momentum tensor \cite{Brown:1992br} and $R$-symmetry current are constructed via the boundary source variational principle for the action \cite{Henningson:1998gx}, \cite{Balasubramanian:1999re}.  The covariant divergence of the energy-momentum tensor has the Brown-York form and yields the diffeomorphism invariance of the renormalized action Ward identity \cite{Imbimbo:1999bj}, thus providing its interpretation as a boundary energy-momentum tensor.  Improvements to both currents are constructed consistent with the Ward identities and trace and $R$-symmetry anomalies.  They are shown to follow from the variation of boundary action improvement terms.

\section{Holographic Supergravity Action \label{section2}}

The bosonic part of the bulk $AdS_5$ ${\cal N}=2$ conformal supergravity action including the single 4 derivative mixed Chern-Simons term is given by
\bea
\Gamma &=& \int d^5 x \det{E} \left[ \frac{1}{2\kappa} R + \Lambda -\frac{Z_A}{4}F_{MN} F^{MN} \right. \cr
 & &\left. \qquad\qquad\qquad\quad + C\, E^{MNRST} F_{MN} F_{RS} A_T - D\, E^{MNRST} R^X_{~~YMN} R^Y_{~~XRS} A_T \right] \cr
 & & \qquad\qquad\qquad\qquad\quad + \frac{1}{\kappa}\int_{\rho = \epsilon} d^4 x \sqrt{-\gamma} K(\gamma) + \Gamma_{\rm Counter-terms},
\eea
where
\be
ds^2 = \frac{R_5^2}{\rho}\left[ g_{\mu\nu} (x,\rho) dx^\mu dx^\nu - \frac{d\rho^2}{4\rho}\right]
\ee
is the Fefferman-Graham metric for the (asymptotic) $AdS_5$ space of radius $R_5$.  Here $\gamma_{\mu\nu}= (R_5^2 /\rho) g_{\mu\nu} (x, \rho)\vert_{\rho=\epsilon}$ is the induced metric on the fixed $\rho = \epsilon$ hypersurface homeomorphic to the $M_4$ boundary.  The extrinsic curvature on that surface is given by the gradient of the normal vector there yielding $K_{\mu\nu} = (\rho / R_5) \partial \gamma_{\mu\nu}/\partial\rho\vert_{\rho=\epsilon}$.  The counter-term boundary action $\Gamma_{\rm Counter-terms}=\int_{\rho = \epsilon} d^4 x \sqrt{-\gamma} B(\gamma)$ is determined through holographic renormalization and normalization conditions as discussed at the end of this section.  The Riemann tensor is given in terms of the Christoffel symbol $\Gamma^R_{MN}$as
\be
R^R_{~~SMN}=\partial_N \Gamma^R_{SM}-\partial_M \Gamma^R_{SN} +\Gamma^L_{SM} \Gamma^R_{LN}- \Gamma^L_{SN} \Gamma^R_{LM} ,
\ee
with $\Gamma^R_{MN}=\frac{1}{2}g^{RS}\left[\partial_N g_{SM}+\partial_M g_{SN}-\partial_S g_{MN}\right]$.  The $U(1)$ $R$-symmetry gauge field $A_M$ has field strength tensor $F_{MN}=\nabla_M A_N -\nabla_N A_M =\partial_M A_N -\partial_N A_M$.  The world tensor $E^{MNRST}=\epsilon^{ABCDE} E_A^{-1M}E_B^{-1N}E_C^{-1R}E_D^{-1S}E_E^{-1T}=(1/\det{E}) \epsilon^{MNRST}$ with the f\"unfbein 
\be
E_M^{~A}=
\begin{pmatrix}
\frac{R_5}{\sqrt{\rho}}e_\mu^{~a} (x,\rho) & 0 \\
 0 & \frac{R_5}{2\rho}
 \end{pmatrix}_{MA}  ,
\ee
where $g_{\mu\nu} (x, \rho) = e_\mu^{~a} (x, \rho) \eta_{ab} e_\nu^{~b} (x, \rho)$ with $\eta_{ab} = (+,-,-,-)_{ab}$.

Varying the bulk action with fixed boundary conditions for the fields yields the field equations.  The Einstein equation has the form
\be
R_{MN}-\frac{1}{2}g_{MN} R=g_{MN} \kappa \Lambda -\kappa T_{MN},
\ee
with the Ricci tensor defined by $R_{MN}\equiv R^R_{~~MNR}$ and the scalar curvature given by $R\equiv g^{MN}R_{MN}$.  The bulk energy-momentum tensor $T_{MN}$ is obtained as
\be
T_{MN} =T^{\rm Maxwell}_{MN} + \nabla_X \Theta^X_{MN} ,
\label{BulkT}
\ee
with the $R-$symmetry gauge field's contribution to the energy-momentum tensor given by the Maxwell symmetric form
\be
T^{\rm Maxwell}_{MN} = Z_A F_{MR} F^R_{~~N} -g_{MN} \left( -\frac{Z_A}{4}F_{RS} F^{RS}\right)
\label{BulkMaxT}
\ee
while the mixed gauge-gravity Chern-Simons contribution to the energy-momentum tensor is
\be
\Theta^X_{MN} = 2D\, E^{QRSTU} F_{TU} \left( g_{QM} R^X_{~~NRS} +g_{QN} R^X_{~~MRS}\right).
\label{BulkMixedCST}
\ee
The covariant derivative is defined according to
\be
\nabla_M T_N^{~~R} \equiv \partial_M T_N^{~~R} -\Gamma^P_{MN} T_P^{~~R} +\Gamma^R_{MP} T_N^{~~P}.
\ee
The Maxwell equation is generalized to include the Chern-Simons terms so that
\be
Z_A \nabla_M F^{ML} +3C\, E^{MNRSL} F_{MN} F_{RS} = D\, E^{MNRSL} R^X_{~~YMN} R^Y_{~~XRS} .
\ee
Using the relations $\nabla_M E_N^{~B}=0=\nabla_M E^{PQRST}$, this can be written as
\be
Z_A\partial_M \left( \det{E}\, F^{ML} \right) +3C \det{E}\,  E^{MNRSL} F_{MN} F_{RS} =D \det{E}\, E^{MNRSL} R^X_{~~YMN} R^Y_{~~XRS} .
\ee

Applying the Fefferman-Graham metric, the Einstein equations can be expanded to have the forms (here the covariant derivative is made utilizing $g_{\mu\nu}$)
\bea
 & & \rho \left[ 2 g^{\prime\prime}_{\mu\nu} - 2 g^\prime_{\mu\lambda} g^{\lambda\rho}g^\prime_{\rho\nu} + g^\prime_{\mu\nu} g^{\rho\sigma}g^\prime_{\rho\sigma}\right] + R_{\mu\nu}(g) -2 g^\prime_{\mu\nu} -g_{\mu\nu} g^{\rho\sigma} g^\prime_{\rho\sigma} \cr
 &=& -\kappa T_{\mu\nu} +\frac{1}{2}g_{\mu\nu} \kappa g^{\rho\sigma} T_{\rho\sigma}
 -\rho g_{\mu\nu} \left[ \frac{1}{3\rho^2}(6+ R_5^2 \kappa \Lambda ) +\frac{1}{6\rho}\kappa g^{\rho\sigma}T_{\rho\sigma} +\frac{4}{3}\kappa T_{44}\right] 
\label{E1}
\eea
\begin{eqnarray}
 g^{\rho\sigma} g^{\prime\prime}_{\rho\sigma} -\frac{1}{2}\left( g^{\rho\sigma} g^\prime_{\sigma\mu} g^{\mu\nu} g^\prime_{\nu\rho} \right)
&  = & \frac{1}{3\rho^2}\left[6 +R^2_5 \kappa \Lambda  \right] +\frac{1}{6\rho}
\kappa g^{\mu\nu} T_{\mu\nu} +\frac{4}{3}\kappa T_{44} 
\label{E2} \\
\nabla_\mu \left( g^{\rho\sigma} g^\prime_{\rho\sigma}\right) - \nabla^\rho g^\prime_{\rho\mu}  & = & 2\kappa T_{\mu 4} ,
\label{E3}
\end{eqnarray}
where the prime indicates differentiation with respect to $\rho$, so that, for example, $g^\prime_{\mu\nu} \equiv \partial g_{\mu\nu}/\partial \rho$.  Likewise the Maxwell equations have the detailed structure (here the permutation tensor $E^{\mu\nu\rho\sigma} (g)=\epsilon^{\mu\nu\rho\sigma} /\sqrt{-g}$ )
\bea
2R_5 Z_A \partial_\mu \left[ \det{e}\, g^{\mu\nu} A^\prime_\nu \right] &+& 3C \det{e}\, E^{\mu\nu\rho\sigma} (g) F_{\mu\nu}(A) F_{\rho\sigma}(A) - D \det{e}\, E^{\mu\nu\rho\sigma}(g) R^\kappa_{~~\lambda\mu\nu}(g) R^\lambda_{~~\kappa\rho\sigma}(g) \cr
 &=& +4\rho D \det{e}\, E^{\mu\nu\rho\sigma}(g) R^{\lambda\kappa}_{~~~~\mu\nu}(g) g^\prime_{\kappa\rho} g^\prime_{\lambda\sigma} \cr
 & &+2\rho D \det{e}\, E^{\mu\nu\rho\sigma}(g) \left( \nabla_\nu g^\prime_{\mu\kappa} -\nabla_\mu g^\prime_{\nu\kappa}\right) g^{\kappa\lambda} \left( \nabla_\sigma g^\prime_{\lambda\rho} -\nabla_\rho g^\prime_{\lambda\sigma}\right)
\label{M1}
\eea
\bea
\rho \frac{\partial}{\partial \rho}\left[ Z_A \det{e}\, g^{\lambda\rho} A^\prime_\rho \right] &-& \frac{1}{4}\partial_\mu \left[Z_A \det{e}\,  g^{\mu\rho} g^{\lambda\sigma} F_{\rho\sigma}(A)\right] = \frac{6\rho}{R_5}C \det{e}\, E^{\tau \rho\sigma\lambda}(g) A^\prime_\tau F_{\rho\sigma}(A) \cr
 & &-\frac{2\rho}{R_5} D \det{e}\, E^{\nu\rho\sigma\lambda}(g) \Biggl[ g^{\mu\tau}\left( -\frac{1}{\rho}g_{\mu\nu} -2\rho g^{\prime\prime}_{\mu\nu} +\rho g^{\pi\kappa} g^\prime_{\kappa\mu} g^\prime_{\pi\nu}\right)\left( \nabla_\sigma g^\prime_{\tau\rho} -\nabla_\rho g^\prime_{\tau\sigma}\right) .\cr
 & & +\left( \frac{1}{2} g^{\tau\pi} \nabla_\nu g^\prime_{\pi\mu} - \Gamma^{\prime \tau}_{\mu\nu}(g)\right) \biggl( R^\mu_{~~\tau\rho\sigma}(g) +\frac{1}{\rho} g_{\tau\rho} \delta_\sigma^\mu -g^\prime_{\tau\rho} \delta_\sigma^\mu -\frac{1}{\rho} g_{\tau\sigma} \delta_\rho^\mu +g^\prime_{\tau\sigma} \delta_\rho^\mu  \cr
 & &\qquad\qquad\qquad\qquad -g_{\tau\rho} g^{\mu\pi} g^\prime_{\pi\sigma} + \rho g^\prime_{\tau\rho} g^{\mu\pi} g^\prime_{\pi\sigma} +g_{\tau\sigma} g^{\mu\pi} g^\prime_{\pi\rho} - \rho g^\prime_{\tau\sigma} g^{\mu\pi} g^\prime_{\pi\rho} \biggr) \Biggr] . 
\label{M2} \nonumber \\
\eea

The boundary currents have an expectation value in the presence of the background gravitational and $U(1)_R$ gauge fields given in terms of the asymptotic behavior of the bulk fields at the $M_4$ boundary of the $AdS_5$ space.  Consequently the field equations need only be solved close to the boundary at $\rho =0$ in order to determine these one point functions.  The $U(1)_R$ gauge field $A_\mu (x, \rho)$ has the asymptotic form in $\rho$ close to the boundary given by
\be
A_\mu (x, \rho) = A_\mu^{(0)} (x) + \rho A_\mu^{(2)} (x) + \rho \ln{(\rho /R_5^2)} ~~ B_\mu (x) .
\ee
Likewise for $\rho \sim 0$ the gravitational field has the behavior
\be
g_{\mu\nu} (x, \rho) = g_{\mu\nu}^{(0)} (x) +\rho g_{\mu\nu}^{(2)} (x) 
 + \rho^2 g_{\mu\nu}^{(4)} (x) + \rho^2 \ln{(\rho /R_5^2)} ~~ h_{\mu\nu} (x).
\ee
Substituting these expansions into the Maxwell equations results in their right hand sides vanishing as $\rho\rightarrow 0$ while the fields $B_\mu$ and the covariant divergence of $A_\mu^{(2)}$ being determined in terms of $A_\mu^{(0)}$.  The transverse part of $A_\mu^{(2)}$ is undetermined.  It corresponds to the other linearly independent solution to the second order Dirichlet problem and it appears as the subleading asymptotic behavior of $A_\mu (x, \rho)$.  A second boundary condition deeper into the bulk would be needed for its specification.  Using $g^{(0)}$ to raise and lower indices the $U(1)_R$ gauge field has the asymptotic solution
\bea
B^\mu (x) &=& \frac{1}{4} \nabla^{(0)}_\nu F^{(0)\nu\mu} \cr
2R_5 Z_A \nabla^{(0)}_\mu A^{(2)\mu} (x) &=& -3C E^{\mu\nu\rho\sigma}(g^{(0)}) F^{(0)}_{\mu\nu} 
F^{(0)}_{\rho\sigma} +D E^{\mu\nu\rho\sigma}(g^{(0)}) R^{(0)\kappa}_{~~~~~\lambda\mu\nu} R^{(0)\lambda}_{~~~~~\kappa\rho\sigma} .
\eea

In order to find the asymptotic solution to the Einstein equations the bulk energy-momentum tensor must be expanded in terms of $\rho$ where it is found from equations (\ref{BulkT}-\ref{BulkMixedCST}) that
\bea
T_{\mu\nu} (x, \rho) &=& \rho T^{(2)}_{\mu\nu} (x) +\cdots \cr
T_{44} (x, \rho) &=& T^{(0)}_{44} (x) +\cdots \cr
T_{\mu 4} (x, \rho) &=& \rho T^{(2)}_{\mu 4} (x) + \rho \ln{(\rho /R_5^2)} \tilde{T}_{\mu 4} (x) +\cdots ,
\eea
with
\bea
T^{(2)}_{\mu\nu} (x) &=& \frac{1}{R^2_5}\left[ Z_A F^{(0)}_{\mu\rho} F^{(0)\rho}_{~~~~~\nu} +\frac{
Z_A}{4} g^{(0)}_{\mu\nu} F^{(0)}_{\rho\sigma} F^{(0)\rho\sigma} \right]\cr
T^{(0)}_{44} (x) &=& -\frac{Z_A}{16 R^2_5} F_{\rho\sigma}^{(0)} F^{(0)\rho\sigma} \cr
T^{(2)}_{\mu 4} (x) &=& -\frac{1}{R^2_5} Z_A F^{(0)}_{\mu\rho} g^{(0)\rho\sigma} \left[ A^{(2)}_\sigma + B_\sigma \right] +\frac{1}{R^3_5} D E^{(0)\rho\sigma\pi\lambda} F^{(0)}_{\pi\lambda} \left[ \nabla^{(0)}_\rho g^{(2)}_{\mu\sigma} - \nabla^{(0)}_\sigma g^{(2)}_{\mu\rho} \right]\cr
 & &-\frac{1}{R^3_5}D E^{(0)\rho\sigma\pi\lambda} \nabla^{(0)}_\alpha \left[ F^{(0)}_{\pi\lambda} \left( R^{(0)\alpha}_{~~~~\mu\rho\sigma} -g^{(0)}_{\mu\rho} g^{(0)\alpha\xi} g^{(2)}_{\xi\sigma} + g^{(0)}_{\mu\sigma} g^{(0)\alpha\xi} g^{(2)}_{\xi\rho} \right)  \right]\cr
\tilde{T}_{\mu 4} (x) &=& -\frac{1}{R^2_5} Z_A F^{(0)}_{\mu\rho} g^{(0)\rho \sigma} B_\sigma  .
\eea
Note that $g^{(0)\rho\sigma} T^{(2)}_{\rho\sigma} =0$ and $\nabla^{(0)\mu}T^{(2)}_{\mu\nu} = 4 \tilde{T}_{\nu 4}$.  It is useful to consider the Maxwell contribution to $T^{(2)}_{\mu 4}$ separately
\be
T^{(2){\rm Maxwell}}_{\mu 4} = -\frac{1}{R^2_5} Z_A F^{(0)}_{\mu\rho} g^{(0)\rho\sigma} \left[ A^{(2)}_\sigma + B_\sigma \right] ,
\ee
and hence the mixed Chern-Simons contribution is just $T^{(2){\rm D}}_{\mu 4} = T^{(2)}_{\mu 4} -T^{(2){\rm Maxwell}}_{\mu 4}$.

The metric coefficients $g^{(2)}$ and $h$ can be determined from the field equation expansion in terms of the boundary metric $g^{(0)}$ and the boundary $R$-symmetry gauge field $A^{(0)}_\mu$ while only the trace and divergence of $g^{(4)}$ is determined in terms of $g^{(0)}$ and $A^{(0)}_\mu$ by the near boundary expansion.  The remaining components of $g^{(4)}$ being fixed by a needed second boundary condition deeper into the bulk for these second order differential equations.  From equation (\ref{E1}) and the vanishing of the bulk energy-momentum tensor $T_{\mu\nu}$ as $\rho \sim 0$ it is found that $\kappa \Lambda R^2_5 = -6$ and 
\be
g^{(2)}_{\mu\nu} = \frac{1}{2}\left[ R^{(0)}_{\mu\nu} -\frac{1}{6} g^{(0)}_{\mu\nu} R^{(0)} \right] ,
\ee
along with $g^{(0)\mu\nu} g^{(2)}_{\mu\nu} = (1/6) R^{(0)}$.  Expanding equation (\ref{E2}) immediately yields $g^{(0)\mu\nu} h_{\mu\nu} =0$ and 
\bea
g^{(0)\mu\nu}g^{(4)}_{\mu\nu} &=& \frac{1}{4}\left( g^{(0)\rho\sigma}g^{(2)}_{\sigma\lambda} g^{(0)\lambda\kappa} g^{(2)}_{\kappa\rho} \right) +\frac{2}{3}\kappa T^{(0)}_{44}\cr
 &=& \frac{1}{16}\left[ R^{(0)}_{\mu\nu} R^{(0)\mu\nu} -\frac{2}{9}R^{(0)2} \right] -\frac{\kappa Z_A}{24 R^2_5} F^{(0)}_{\mu\nu} F^{(0)\mu\nu} .\cr
 & & 
\eea
Returning to equation (\ref{E1}) $h_{\mu\nu}$ is obtained
\bea
h_{\mu\nu} &=& \frac{1}{2} g^{(2)}_{\mu\lambda} g^{(0)\lambda\rho} g^{(2)}_{\rho\nu} -\frac{1}{8}g^{(0)}_{\mu\nu} \left[g^{(0)\rho\sigma}g^{(2)}_{\sigma\lambda} g^{(0)\lambda\kappa} g^{(2)}_{\kappa\rho}\right] -\frac{1}{4} R^{(2)}_{\mu\nu} -\frac{1}{4} \kappa T^{(2)}_{\mu\nu} \cr
 &=& \frac{1}{8} \left[ R^{(0)}_{\mu\rho} R^{(0)\rho}_{~~~~\nu} -\frac{1}{3} R^{(0)} R^{(0)}_{\mu\nu} -\frac{1}{36} g^{(0)}_{\mu\nu} R^{(0)2} \right] -\frac{1}{32} g^{(0)}_{\mu\nu} \left[ R^{(0)}_{\mu\nu} R^{(0)\mu\nu} -\frac{2}{9}R^{(0)2}  \right] \cr
 & &-\frac{1}{8}\left[ R^{(0)}_{\mu\rho} R^{(0)\rho}_{~~~\nu} - R^{(0)\rho\sigma} R^{(0)}_{\sigma\mu\nu\rho} + \frac{1}{6} \nabla^{(0)}_\mu \nabla^{(0)}_\nu R^{(0)} -\frac{1}{2} \nabla^{(0)}_\rho \nabla^{(0)\rho} R^{(0)}_{\mu\nu} +\frac{1}{12} g^{(0)}_{\mu\nu} \nabla^{(0)}_\rho \nabla^{(0)\rho} R^{(0)} \right] \cr
 & &-\frac{1}{4R^2_5} \kappa \left[ Z_A F^{(0)}_{\mu\rho} F^{(0)\rho}_{~~~~~\nu} +\frac{
Z_A}{4} g^{(0)}_{\mu\nu} F^{(0)}_{\rho\sigma} F^{(0)\rho\sigma} \right] ,
\eea
where the Ricci tensor has been expanded close to the boundary as $R_{\mu\nu} (g) = R^{(0)}_{\mu\nu} (g^{(0)}) + \rho R^{(2)}_{\mu\nu} (g^{(0)}) + \cdots$ and likewise for the scalar curvature
$R(g)= R^{(0)} (g^{(0)}) + \rho R^{(2)} (g^{(0)}) +\cdots$.  The expansion coefficients are given by
\bea
R^{(2)}_{\mu\nu} (g^{(0)}) &=& \frac{1}{2}\left[ R^{(0)}_{\mu\rho} R^{(0)\rho}_{~~~\nu} - R^{(0)\rho\sigma} R^{(0)}_{\sigma\mu\nu\rho} + \frac{1}{6} \nabla^{(0)}_\mu \nabla^{(0)}_\nu R^{(0)} -\frac{1}{2} \nabla^{(0)}_\rho \nabla^{(0)\rho} R^{(0)}_{\mu\nu} +\frac{1}{12} g^{(0)}_{\mu\nu} \nabla^{(0)}_\rho \nabla^{(0)\rho} R^{(0)} \right]\cr
R^{(2)} (g^{(0)}) &=& -\frac{1}{2}\left[ R^{(0)}_{\mu\nu} R^{(0)\mu\nu} -\frac{1}{6} R^{(0)2}\right].
\eea
Note that $g^{(0)\mu\nu} h_{\mu\nu} =0$ as found above.  Finally equation (\ref{E3}) yields the divergence of the coefficients $g^{(2)}$, $h$ and $g^{(4)}$
\bea
\nabla^{(0)\rho} g^{(2)}_{\rho\mu} &=& \frac{1}{6}\nabla^{(0)}_\mu R^{(0)}\cr
\nabla^{(0)\rho} h_{\rho\mu} &=& -\kappa \tilde{T}_{\mu4} = \frac{\kappa Z_A}{R^2_5} F^{(0)}_{\mu\rho} g^{(0)\rho\sigma} B_\sigma =-\frac{1}{4}\nabla^{(0)\rho} T^{(2)}_{\rho\mu} \cr
\nabla^{(0)\rho} g^{(4)}_{\rho\mu} &=& -\frac{1}{2}\nabla^{(0)\rho} h_{\rho\mu}+\frac{1}{2} 
\nabla^{(0)\rho} \left[g^{(2)}_{\rho\sigma} g^{(0)\sigma\lambda} g^{(2)}_{\lambda\mu}\right]\cr
 & &-\frac{1}{4}\nabla^{(0)\rho}\left[g^{(2)}_{\rho\mu} \left( g^{(0)\alpha\beta} g^{(2)}_{\alpha\beta}\right) \right] +\frac{1}{8}\nabla^{(0)\rho}\left[g^{(0)}_{\rho\mu} \left( g^{(0)\alpha\beta} g^{(2)}_{\alpha\beta}\right)^2 \right] \cr
 & &-\frac{1}{8}\nabla^{(0)\rho}\left[g^{(0)}_{\rho\mu} \left( g^{(0)\xi\sigma} g^{(2)}_{\sigma\kappa}g^{(0)\kappa\lambda} g^{(2)}_{\lambda\xi}\right)\right] -\kappa T^{(2)}_{\mu 4} +\frac{2}{3} \kappa \nabla^{(0)}_\mu T^{(0)}_{44} .
\eea

The near-boundary analysis of the field equations allows the boundary divergences of the action, now regulated at the surface $\rho = \epsilon$, to be determined.  The regulated action is given by
\bea
\Gamma_{\rm Reg.} &=& \int d^4 x \int_{\rho = \epsilon} d\rho \det{E} \left[ \frac{1}{2\kappa} R + \Lambda -\frac{Z_A}{4}F_{MN} F^{MN} \right. \cr
 & &\left. \qquad\qquad\qquad\qquad\quad + C\, E^{MNRST} F_{MN} F_{RS} A_T - D\, E^{MNRST} R^X_{~~YMN} R^Y_{~~XRS} A_T \right] \cr
 & & \qquad\qquad\qquad\qquad\qquad\qquad\quad + \frac{1}{\kappa}\int_{\rho = \epsilon} d^4 x \sqrt{-\gamma} K(\gamma) .
\eea
Applying the Fefferman-Graham form of the metric and employing the near-boundary solutions to the field equations the divergent terms in the action are isolated as
\bea
\Gamma_{\rm Reg.} &=& \int d^4 x \int_{\rho = \epsilon} d\rho  \sqrt{-g^{(0)}} \left\{ -\frac{1}{\rho^3}\frac{6R^3_5}{\kappa}\right. \cr
 & &\left. \qquad\qquad\qquad -\frac{1}{\rho}\frac{R^3_5}{16\kappa}\left[\left( R^{(0)}_{\mu\nu} R^{(0)\mu\nu} -\frac{1}{3}R^{(0)2}\right) +\frac{2\kappa Z_A}{R^2_5}F^{(0)}_{\mu\nu} F^{(0)\mu\nu} \right] +\ldots \right\} \cr
 &=&\int d^4 x \sqrt{-g^{(0)}} \left\{ -\frac{1}{\epsilon^2}\frac{3R^3_5}{\kappa} \right. \cr
 & &\left. +\ln{(\epsilon /R_5^2)}\frac{R^3_5}{16\kappa}\left[\left( R^{(0)}_{\mu\nu} R^{(0)\mu\nu} -\frac{1}{3}R^{(0)2}\right) +\frac{2\kappa Z_A}{R^2_5}F^{(0)}_{\mu\nu} F^{(0)\mu\nu} \right] \right\} +O(\epsilon^0).
\eea
The holographically renormalized action is defined by choosing the near-boundary counter-term action $\Gamma_{\rm Counter-terms}$ to cancel the divergent terms in the regulated action and to impose normalization conditions on the remaining finite terms so that
\be
\Gamma_{\rm Sub.} = \Gamma_{\rm Reg.} + \Gamma_{\rm Counter-terms} ,
\ee
where, after inverting the near-boundary expansion of the fields to write the boundary quantities in terms of the tensors at the surface $\rho = \epsilon$, the near-boundary counter-term action is found to be
\bea
\Gamma_{\rm Ct.-terms} &=& \frac{3}{\kappa R_5}\int_{\rho = \epsilon} d^4 x \sqrt{-\gamma}  -\frac{R_5}{4\kappa}\int_{\rho = \epsilon} d^4 x \sqrt{-\gamma} R(\gamma) \cr
 & &-\ln{(\epsilon /R_5^2)}\frac{R^3_5}{16\kappa}\int_{\rho = \epsilon} d^4 x \sqrt{-\gamma}\left[\left( R_{\mu\nu}(\gamma) R^{\mu\nu}(\gamma) -\frac{1}{3}R^{2}(\gamma)\right) +\frac{2\kappa Z_A}{R^2_5}F_{\mu\nu}(A) F^{\mu\nu}(A) \right] ,\cr
 & & 
\eea
where the induced metric $\gamma_{\mu\nu} = (R^2_5/\rho) g_{\mu\nu}\vert_{\rho=\epsilon}$ on the near-boundary suface is used in the counter-term action.  The finite holographic normalization is chosen through the dimensionless ratio in the logarithmic counter-term as $(\epsilon /R_5^2)$.  A different normalization such as $(\epsilon /\tau R_5^2 )$, with $\tau \in {\mathbb R}^+$, will correspond to a finite boundary term in the action,\
\be
\Gamma_{\rm Improve}= \ln{\tau}\frac{R^3_5}{16\kappa}\int_{\rho = \epsilon} d^4 x \sqrt{-\gamma}\left[\left( R_{\mu\nu}(\gamma) R^{\mu\nu}(\gamma) -\frac{1}{3}R^{2}(\gamma)\right) +\frac{2\kappa Z_A}{R^2_5}F_{\mu\nu}(A) F^{\mu\nu}(A) \right] ,
\label{ImprovementCounter-term}
\ee
that will lead to finite holographic improvements to the currents that do not alter the form of the scale and chiral $R$ anomalies and are consistent with the current Ward identities as discussed in the next section.


\section{Holographic Currents \label{section3}}

According to gravity/gauge duality, the expectation values of the boundary currents, the energy-momentum tensor $\theta_{\mu\nu} (x)$ and the $R$-symmetry current $R_\mu (x)$, in the presence of their respective external sources $g^{(0)}_{\mu\nu}(x)$ and $A^{(0)}_\mu (x)$ are found by varying the boundary sources in the on-shell renormalized action as
\be
\delta\Gamma [g^{(0)}_{\mu\nu}, A^{(0)}_\mu ]= \int d^4 x \sqrt{-g^{(0)}}\left[ \frac{1}{2} \theta_{\mu\nu}(x) \delta g^{(0)\mu\nu} (x) + R_\mu (x) \delta A^{(0)\mu} (x) \right].
\label{Currents}
\ee
This can be accomplished by varying the sources in the regulated near boundary action and counter-terms, then taking the $\epsilon \rightarrow 0$ limit of the subtracted action.  The variation of the surface $\rho = \epsilon$ sources $\gamma$ and $A$ for the on-shell, near boundary regulated Einstein-Maxwell-Chern-Simons action (i.e. the second order in derivative part of the action, the mixed gauge-gravity Chern-Simons term has higher order derivatives and will be treated separately) yields
\bea
\delta \Gamma_{\rm Reg.}^{EMCS} &=& \int_{\rho =\epsilon} d^5 x \det{E} \biggl\{ \frac{1}{2\kappa} \nabla_R \left( \nabla^R g_{MN}\delta g^{MN} -\nabla_S \delta g^{RS} \right) \cr
 & &+ \nabla_M \left[ \delta A_L \left( -Z_A F^{ML} +8 C E^{MLRST} \partial_R 
A_S A_T \right)  \right] \biggr\} \cr
 &=&-\frac{2}{\kappa} \int_{\rho =\epsilon} d^4 x \frac{\rho}{R_5} \frac{\partial}{\partial \rho} \left[ \delta \sqrt{-\gamma}\right] + \frac{1}{2\kappa} \int_{\rho =\epsilon} d^4 x \sqrt{-\gamma} \left[ K_{\mu\nu} (\gamma) -\gamma_{\mu\nu} K(\gamma) \right] \delta \gamma^{\mu\nu} \cr
 & & -\int_{\rho =\epsilon} d^4 x \sqrt{-\gamma} \delta A_\mu \left[ 2Z_A \frac{\rho}{R_5} \gamma^{\mu\nu} \frac{\partial}{\partial \rho}A_\nu + 8C E^{\mu\nu\rho\sigma} (\gamma) A_\nu \partial_\rho A_\sigma \right].
\eea

In order for there to be a well defined variational problem the extrinsic curvature term must be added to cancel the undetermined source variation $\partial/\partial \rho (\delta \gamma^{\mu\nu})$ terms.  The variation of the near boundary extrinsic curvature term, $\Gamma_{\rm K}= 1/\kappa \int_\epsilon d^4 x \sqrt{-\gamma} K(\gamma)$, is found to be
\be
\delta \Gamma_{\rm K} = \frac{2}{\kappa} \int_\epsilon d^4 x \frac{\rho}{R_5} \frac{\partial}{\partial \rho} \left( \delta \sqrt{\gamma}\right).
\ee
In addition, the variation of the counter-terms,
\bea
\delta\Gamma_{\rm Ct.-Terms} &=& -\frac{3}{2\kappa R_5} \int_\epsilon d^4 x \sqrt{-\gamma} \gamma_{\mu\nu} \delta\gamma^{\mu\nu} -\frac{R_5}{2\kappa} \int_\epsilon d^4 x \sqrt{-\gamma} \frac{1}{2} \left[R_{\mu\nu} (\gamma) -\frac{1}{2}\gamma_{\mu\nu} R(\gamma) \right]\delta \gamma^{\mu\nu} \cr
 & &\qquad\qquad -R_5 Z_A \ln{\frac{\epsilon}{R_5^2}} \int_\epsilon d^4 x \sqrt{-\gamma}\delta\gamma^{\mu\nu} \left[ F_{\mu\rho} F_{\nu\sigma} \gamma^{\rho\sigma}  -\frac{1}{4} \gamma_{\mu\nu} F^2 \right]\cr
 & &\qquad\qquad\qquad + R_5 Z_A \ln{\frac{\epsilon}{R_5^2}} \int_\epsilon d^4 x \sqrt{-\gamma} \delta A_\sigma \nabla_\rho F^{\rho\sigma} ,
\eea
are needed to provide a finite $\epsilon \rightarrow 0$ limit for the Einstein and Maxwell terms contributions to the currents as well as the action (both Chern-Simons terms are finite as are their contributions to the currents).

Combining these terms, $\Gamma_{\rm Sub.}^{EMCS}= \Gamma_{\rm Reg.}^{EMCS} +\Gamma_{\rm K} +\Gamma_{\rm Ct.-Terms}$, it is found that the subtracted action is given by
\bea
\delta \Gamma_{\rm Sub.} &=& \frac{\epsilon}{R_5^2} \int_\epsilon d^4 x \sqrt{-\gamma} \delta\gamma^{\mu\nu} \frac{1}{2} \left[ \frac{R_5^2}{\epsilon \kappa}\left( K_{\mu\nu} (\gamma) -K(\gamma ) \gamma_{\mu\nu} \right) -\frac{3R_5}{\epsilon\kappa} \gamma_{\mu\nu} -\frac{R_5^3}{2\epsilon\kappa} \left( R_{\mu\nu} (\gamma) -\frac{1}{2} R(\gamma) \gamma_{\mu\nu} \right) \right. \cr
 & &\left. \qquad\qquad\qquad\qquad\qquad\qquad  -2R_5 Z_A \ln{\frac{\epsilon}{R_5^2}} \frac{R^2_5}{\epsilon} \left( F_{\mu\rho} F_{\nu\sigma} \gamma^{\rho\sigma} -\frac{1}{4} \gamma_{\mu\nu} F^2 \right) \right] \cr
 & &-\int_{\rho =\epsilon} d^4 x \sqrt{-\gamma} \delta A_\mu \left[ 2Z_A \frac{\rho}{R_5} \gamma^{\mu\nu} \frac{\partial}{\partial \rho}A_\nu + 8C E^{\mu\nu\rho\sigma} (\gamma) A_\nu \partial_\rho A_\sigma  - R_5 Z_A \ln{\left(\frac{\epsilon}{R_5^2}\right)}  \nabla_\rho F^{\rho\mu}\right] . \cr
 & & 
\eea
Thus the subtracted boundary energy-momentum tensor and $R$-symmetry current take the form
\bea
\theta_{{\rm Sub.}\mu\nu}^{\rm EMCS} (\epsilon)&\equiv& \frac{R_5^2}{\epsilon} \frac{2}{\sqrt{-\gamma}} \frac{\delta \Gamma_{\rm Sub.}}{\delta \gamma^{\mu\nu}} \cr
 &=&\frac{R_5^2}{\epsilon\kappa} \left( K_{\mu\nu} (\gamma) -K(\gamma ) \gamma_{\mu\nu} \right) -\frac{3R_5}{\epsilon\kappa} \gamma_{\mu\nu} -\frac{R_5^3}{2\epsilon\kappa} \left( R_{\mu\nu} (\gamma) -\frac{1}{2} R(\gamma) \gamma_{\mu\nu} \right) \cr
 & &-2R_5 Z_A \ln{\frac{\epsilon}{R_5^2}} \frac{R^2_5}{\epsilon} \left( F_{\mu\rho} F_{\nu\sigma} \gamma^{\rho\sigma} -\frac{1}{4} \gamma_{\mu\nu} F^2 \right) \cr
R_{\rm Sub.}^{{\rm EMCS}\mu} (\epsilon) &\equiv&   \frac{1}{\sqrt{-\gamma}} \frac{\delta \Gamma_{\rm Sub.}}{\delta A_{\mu}} \cr
 &=&-2Z_A \gamma^{\mu\nu} \frac{\epsilon}{R_5} \frac{\partial}{\partial \rho} A_\nu  +8CE^{\mu\nu\rho\sigma} (\gamma) A_\nu \partial_\rho A_\sigma + R_5 Z_A \ln{\left(\frac{\epsilon}{R_5^2}\right)} \nabla_\rho F^{\rho\sigma} .
\eea

Expanding these expressions as the $\rho =0$ boundary is approached gives the Einstein-Maxwell-pure $U(1)_R$ gauge field Chern-Simons contribution to the renormalized boundary currents as
\bea
\theta_{\mu\nu}^{\rm EMCS} &=&\theta_{{\rm Sub.}\mu\nu}^{\rm EMCS} (\epsilon =0)= \frac{R_5^3}{\kappa} \left[ 2 g^{(4)}_{\mu\nu} +h_{\mu\nu}- g^{(2)}_{\mu\nu} g^{(0)\rho\sigma} g^{(2)}_{\rho\sigma} \right. \cr
 & &\left. \qquad\qquad\qquad\qquad -g^{(0)}_{\mu\nu} \left( 2 g^{(0)\rho\sigma} g^{(4)}_{\rho\sigma} + g^{(2)\rho\sigma} g^{(2)}_{\rho\sigma}\right) \right]\cr
 & &-\frac{R_5^3}{4\kappa} \left[ R^{(0)}_{\mu\lambda} R^{(0)\lambda}_{~~~\nu} +R^{(0)}_{\lambda\mu\rho\nu} R^{(0)\lambda\rho} -\frac{1}{2} \nabla^{(0)2} R^{(0)}_{\mu\nu} +\frac{1}{6}\nabla^{(0)}_\mu  \nabla^{(0)}_\nu R^{(0)} \right. \cr
 & &\left. \qquad\qquad\qquad\qquad +\frac{1}{12} g^{(0)}_{\mu\nu}\nabla^{(0)2} R^{(0)2}  \right.  \cr
 & &\left. +\frac{1}{2} R^{(0)}_{\rho\sigma} R^{(0)\rho\sigma} g^{(0)}_{\mu\nu} -\frac{1}{12} g^{(0)}_{\mu\nu}  R^{(0)2} -\frac{1}{2} R^{(0)}\left(R^{(0)}_{\mu\nu} -\frac{1}{6} g^{(0)}_{\mu\nu} R^{(0)} \right)   \right] \cr
 & & \cr
R^{{\rm EMCS}\mu} &=& R_{\rm Sub.}^{{\rm EMCS}\mu} (\epsilon =0) =  -2R_5 Z_A \left( A^{(2)\mu} + B^\mu \right) -4C E^{\mu\nu\rho\sigma} (g^{(0)}) A^{(0)}_\nu F^{(0)}_{\rho\sigma} .
\eea
These are precisely the same expressions obtained by expanding the variation of the action terms $\delta\Gamma^{\rm EMCS}=\lim_{\epsilon\rightarrow 0}\delta\Gamma_{\rm Sub.}$ at the boundary directly with
\bea
\theta_{\mu\nu}^{\rm EMCS} &\equiv& \frac{2}{\sqrt{-g^{(0)}}} \frac{\delta \Gamma^{\rm EMCS}}{\delta g^{(0)\mu\nu}} \cr
R^{{\rm EMCS}\mu} &\equiv&  \frac{1}{\sqrt{-g^{(0)}}} \frac{\delta \Gamma^{\rm EMCS}}{\delta A^{(0)}_{\mu}}  . 
\eea

Finally the contribution to the currents due to the higher derivative mixed gauge-gravity Chern-Simons term can be determined directly from the variation of its action at the boundary as it is finite
\bea
\delta \Gamma_{\rm RRA}&=& -4D \int d^5 x \partial_P \left[ \det{E} E^{PQRSL} g_{QM} A_L R^X_{~~NRS} \nabla_X \delta g^{MN}  \right] \cr
 & &+4D \int d^5 x \partial_X \left[ \det{E} E^{PQRSL} g_{QM} \delta g^{MN} \nabla_P\left(A_L R^X_{~~NRS} \right)\right] .
\eea
The second term when expanded in terms of the Fefferman-Graham metric can be seen to vanish at the $\rho=0$ boundary:
\bea
\delta \Gamma_{\rm RRA}^{\rm second~ term} &=& -2D \int d^4 x \delta g^{\mu\nu} \sqrt{-g} \left[  E^{\xi\sigma\pi\lambda} (g) F_{\pi\lambda} \left\{ \rho g_{\xi\mu} \left( -2 g^{\prime\prime}_{\sigma\nu} +g^\prime_{\nu\rho} g^{\rho\zeta} g^\prime_{\zeta\sigma} \right) \right.\right. \cr
 & &\left.\left. \qquad\qquad\qquad\qquad\qquad\qquad\qquad\qquad +\rho g_{\xi\nu} \left( -2 g^{\prime\prime}_{\sigma\mu} +g^\prime_{\mu\rho} g^{\rho\zeta} g^\prime_{\zeta\sigma} \right)\right\} \right. \cr
 & &\left. +\rho E^{\rho\sigma\xi\lambda} (g) A^\prime_\lambda \left\{ g_{\xi\mu} \left( \nabla_\sigma g^\prime_{\nu\rho} -\nabla_\rho g^\prime_{\nu\sigma} \right) + g_{\xi\nu} \left( \nabla_\sigma g^\prime_{\mu\rho} -\nabla_\rho g^\prime_{\mu\sigma} \right) \right\}\right] \cr
 & & \cr
 & &\longrightarrow  0 ~~~~{\rm as}~~~{\rho \rightarrow 0}  .
\eea
The first expression for the variation of the mixed Chern-Simons action for finite $\rho = \epsilon$ has an undetermined $\partial /\partial \rho \delta \gamma^{\mu\nu}$ term in it.  However this contribution vanishes on the $\rho =0$ boundary contrary to the extrinsic curvature case discussed earlier
\bea
\delta \Gamma_{\rm RRA}^{\rm first~ term} &=& -4D \int d^4 x \sqrt{-g^{(0)}} \frac{1}{2} \delta g^{(0)\mu\nu} E^{\xi\rho\sigma\lambda} (g^{(0)}) \left\{ \nabla^{(0)}_\alpha \left[ A_\lambda^{(0)} \left( g^{(0)}_{\xi\mu} R^{(0)\alpha}_{~~~\nu\rho\sigma} + g^{(0)}_{\xi\nu} R^{(0)\alpha}_{~~~\mu\rho\sigma}\right) \right] \right.  \cr
 & &\left. +2\nabla^{(0)}_\rho \left[A^{(0)}_\lambda \left(g^{(0)}_{\xi\mu} g^{(2)}_{\nu\sigma} + g^{(0)}_{\xi\nu} g^{(2)}_{\mu\sigma} \right) \right] 
 -2 A^{(0)}_\lambda \left(g^{(0)}_{\xi\mu}\nabla^{(0)}_\rho g^{(2)}_{\nu\sigma} + g^{(0)}_{\xi\nu} \nabla^{(0)}_\rho g^{(2)}_{\mu\sigma} \right)\right\} . 
\eea
This leads to the final mixed gravitational and $U(1)_R$ gauge field Chern-Simons contribution to the energy-momentum tensor of the form
\bea
\theta^{\rm CS}_{\mu\nu} &=& \frac{2}{\sqrt{-g^{(0)}}} \frac{\delta \Gamma_{\rm RRA}}{\delta g^{(0)\mu\nu}} 
 = -4D E^{\xi\rho\sigma\lambda} (g^{(0)}) \left\{ \nabla^{(0)}_\alpha \left[ A_\lambda^{(0)} \left( g^{(0)}_{\xi\mu} R^{(0)\alpha}_{~~~\nu\rho\sigma} + g^{(0)}_{\xi\nu} R^{(0)\alpha}_{~~~\mu\rho\sigma}\right) \right] \right.  \cr
 & &\left. +2\nabla^{(0)}_\rho \left[A^{(0)}_\lambda \left(g^{(0)}_{\xi\mu} g^{(2)}_{\nu\sigma} + g^{(0)}_{\xi\nu} g^{(2)}_{\mu\sigma} \right) \right] 
 -2 A^{(0)}_\lambda \left(g^{(0)}_{\xi\mu}\nabla^{(0)}_\rho g^{(2)}_{\nu\sigma} + g^{(0)}_{\xi\nu} \nabla^{(0)}_\rho g^{(2)}_{\mu\sigma} \right)\right\} .
\eea

Thus the complete renormalized boundary energy-momentum tensor and $R$-symmetry current are determined as
\bea
\theta_{\mu\nu} &=& \theta_{\mu\nu}^{\rm EMCS}  +\theta^{\rm CS}_{\mu\nu} \cr
 &=& \frac{R_5^3}{\kappa} \left[ 2 g^{(4)}_{\mu\nu} +h_{\mu\nu}- g^{(2)}_{\mu\nu} g^{(0)\rho\sigma} g^{(2)}_{\rho\sigma} -g^{(0)}_{\mu\nu} \left( 2 g^{(0)\rho\sigma} g^{(4)}_{\rho\sigma} + g^{(2)\rho\sigma} g^{(2)}_{\rho\sigma}\right) \right]\cr
 & &-\frac{R_5^3}{4\kappa} \left[ R^{(0)}_{\mu\lambda} R^{(0)\lambda}_{~~~~\nu} +R^{(0)}_{\lambda\mu\rho\nu} R^{(0)\lambda\rho} -\frac{1}{2} \nabla^{(0)2} R^{(0)}_{\mu\nu} +\frac{1}{6}\nabla^{(0)}_\mu  \nabla^{(0)}_\nu R^{(0)} +\frac{1}{12} g^{(0)}_{\mu\nu}\nabla^{(0)2} R^{(0)2}  \right.  \cr
 & &\left. +\frac{1}{2} R^{(0)}_{\rho\sigma} R^{(0)\rho\sigma} g^{(0)}_{\mu\nu}  -\frac{1}{2} R^{(0)}R^{(0)}_{\mu\nu} \right] \cr
 & & -4D E^{\xi\rho\sigma\lambda} (g^{(0)}) \left\{ \nabla^{(0)}_\alpha \left[ A_\lambda^{(0)} \left( g^{(0)}_{\xi\mu} R^{(0)\alpha}_{~~~\nu\rho\sigma} + g^{(0)}_{\xi\nu} R^{(0)\alpha}_{~~~\mu\rho\sigma}\right) \right] \right.  \cr
 & &\left. +2\nabla^{(0)}_\rho \left[A^{(0)}_\lambda \left(g^{(0)}_{\xi\mu} g^{(2)}_{\nu\sigma} + g^{(0)}_{\xi\nu} g^{(2)}_{\mu\sigma} \right) \right] 
 -2 A^{(0)}_\lambda \left(g^{(0)}_{\xi\mu}\nabla^{(0)}_\rho g^{(2)}_{\nu\sigma} + g^{(0)}_{\xi\nu} \nabla^{(0)}_\rho g^{(2)}_{\mu\sigma} \right)\right\} \cr
 & & \cr
R^\mu &=& R^{{\rm EMCS}\mu} \cr
 &=& -2R_5 Z_A \left( A^{(2)\mu} + B^\mu \right) -4C E^{\mu\nu\rho\sigma} (g^{(0)}) A^{(0)}_\nu F^{(0)}_{\rho\sigma} .
\label{finalcurrents}
\eea

Exploiting the near boundary solutions to the field equations found in section \ref{section2} the anomalous divergence of the $R$-symmetry current is found to be
\be
\nabla_\mu R^\mu = C E^{(0)\mu\nu\rho\sigma} F^{(0)}_{\mu\nu} F^{(0)}_{\rho\sigma} + D E^{(0)\mu\nu\rho\sigma} R^{(0)\xi\zeta}_{\mu\nu} R^{(0)}_{\xi\zeta\rho\sigma}  .
\ee
The energy-momentum tensor has contributions from gravity and matter flowing into the boundary of the form
\bea
\nabla^\mu \theta^{\rm EMCS}_{\mu\nu} &=&-2R_5^3 T^{(2)}_{\nu 4} = -2R_5^3 \left( T^{(2){\rm Maxwell}}_{\nu 4} +T^{(2){\rm D}}_{\nu 4} \right) \cr
\nabla^\mu \theta^{\rm CS}_{\mu\nu} &=& 2 R_5^3 T^{(2)\rm D}_{\nu 4} + A^{(0)}_\nu {\cal A}^{\rm D}_{R} ,
\eea
with the mixed Chern-Simons contribution to the $R$-symmetry anomaly given by
\be
{\cal A}^{\rm D}_{R} = D E^{\mu\nu\rho\sigma} (g^{(0)}) R^{(0)\xi\zeta}_{~~~~~\mu\nu} R^{(0)}_{\rho\sigma\xi\zeta} .
\ee
From these follows the Ward identity relating the diverence of the energy-momentum tensor with that of the $R$-symmetry current as
\be
\nabla^\mu \theta_{\mu\nu} = F^{(0)}_{\mu\nu} R^\mu + A^{(0)}_\nu \nabla_\mu R^\mu .
\label{WardIdentity}
\ee

The trace (taken with $g^{(0)\mu\nu}$) for the various contributions to the energy-momentum tensor is found from the field equations to be
\bea
\theta^{\rm EMCS~\mu}_{~~~~\mu} &=& \frac{R_5^3}{\kappa} \left[ \frac{1}{8}\left( R^{(0)}_{\mu\nu} R^{(0)\mu\nu} - \frac{1}{3}R^{(0)2} \right)+ \frac{\kappa Z_A}{4 R_5^2} F^{(0)}_{\mu\nu} F^{(0)\mu\nu} \right] \cr
 &\equiv & {\cal A} ,
\eea
while
\be
\theta^{\rm CS~\mu}_{~~\mu} = 0.
\ee
Thus the renomalized boundary energy-momentum tensor has the anomalous trace 
\be
\theta^{~\mu}_{\mu} = {\cal A} .
\label{TraceAnomaly}
\ee

These results agree with the general diffeomorphism and $R$-symmetry transformations of the action.  From the definition of the boundary currents, equation (\ref{Currents}), it is found that the action is invariant under diffeomorphism transformations
\bea
\delta g^{(0)\mu\nu} &=& \nabla^\mu \xi^\nu +\nabla^\nu \xi^\mu  \cr
\delta A^{(0)\mu} &=& -\nabla^\mu \xi_\nu A^{(0)\nu} -\xi_\nu \nabla^\nu A^{(0)\mu} ,
\label{diffeo}
\eea
so that
\bea
\delta\Gamma &=& \int d^4 x \sqrt{-g^{(0)}} \xi^\nu \left[ -\nabla^\mu \theta_{\mu\nu} +F^{(0)}_{\mu\nu} R^\mu + A^{(0)}_\nu \nabla_\mu R^\mu \right] \cr
 &=& 0 ,
\label{diffeo2}
\eea
as follows from equation (\ref{WardIdentity}).  The anomalous $R$-symmetry variation of the action follows directly from the $R$-current divergence equation.  For the $R$-symmetry transformations
\be
\delta g^{(0)\mu\nu} =0 \quad , \quad \delta A^{(0)\mu} = \partial^\mu \omega ,
\label{Rvar}
\ee
equation (\ref{Currents}) implies
\bea
\delta\Gamma &=&-\int d^4 x \sqrt{-g^{(0)}} \omega \nabla^{(0)\mu} R_\mu \cr
 &=&-\int d^4 x \sqrt{-g^{(0)}} \omega \left[ C E^{(0)\mu\nu\rho\sigma} F^{(0)}_{\mu\nu} F^{(0)}_{\rho\sigma} + D E^{(0)\mu\nu\rho\sigma} R^{(0)\xi\zeta}_{\mu\nu} R^{(0)}_{\xi\zeta\rho\sigma}  \right]  .
\label{Rvar2}
\eea
The energy-momentum trace anomaly equation (\ref{TraceAnomaly}) implies the Weyl scale transformation,
\be
\delta g^{(0)\mu\nu} =2 \sigma g^{(0)\mu\nu} \quad , \quad \delta A^{(0)\mu} = 0 ,
\label{Weyl1}
\ee
anomaly for the renormalized action
\bea
\delta\Gamma &=& \int d^4 x \sqrt{-g^{(0)}} \sigma (x) \theta^\mu_{~~\mu} (x) = \int d^4 x \sqrt{-g^{(0)}} \sigma (x) {\cal A} (x) \cr
 &=&\int d^4 x \sqrt{-g^{(0)}} \sigma (x) \frac{R_5^3}{\kappa} \left[ \frac{1}{8}\left( R^{(0)}_{\mu\nu} R^{(0)\mu\nu} - \frac{1}{3}R^{(0)2} \right)+ \frac{\kappa Z_A}{4 R_5^2} F^{(0)}_{\mu\nu} F^{(0)\mu\nu} \right] .
\label{Weyl2}
\eea

It is possible to improve the definition of the currents by adding a finite boundary counter-term to the action given by equation (\ref{ImprovementCounter-term}) which at the $\rho =0$ boundary becomes
\be
\Gamma_{\rm Improve}= \ln{\tau}\frac{R^3_5}{2\kappa}\int d^4 x \sqrt{-g^{(0)}}\left[\frac{1}{8}\left( R_{\mu\nu}^{(0)} R^{(0)\mu\nu} -\frac{1}{3}R^{(0)2}\right) +\frac{\kappa Z_A}{4R^2_5}F_{\mu\nu}^{(0)} F^{(0)\mu\nu}\right].
\label{ImprovementCounter-term2}
\ee
Its variation at the boundary is given by
\bea
\delta \Gamma_{\rm Improve} &=& \ln{\tau} \frac{R_5^3}{2\kappa} \int d^4 x \sqrt{-g^{(0)}} \frac{1}{2} \delta g^{(0)\mu\nu} 4h_{\mu\nu} \cr
 & &-\ln{\tau} \frac{R_5^3}{2\kappa} \int d^4 x \sqrt{-g^{(0)}} \delta A_\nu^{(0)}\frac{\kappa Z_A}{R_5^2} \nabla_\mu^{(0)} F^{(0)\mu\nu} .
\eea
These variations lead to the energy-momentum tensor, $\tau_{\mu\nu}$, and $R$-symmetry current, $r^\mu$, improvement terms
\bea
\tau_{\mu\nu} &=& \frac{R_5^3}{2\kappa} \ln{\tau} \,4 h_{\mu\nu} \cr
r^\mu &=& -\frac{R_5^3}{2\kappa} \ln{\tau} \, \frac{\kappa Z_A}{R_5^2} \nabla_\nu^{(0)} F^{(0)\nu\mu} = -\frac{R_5^3}{2\kappa} \ln{\tau} \,4\frac{\kappa Z_A}{R_5^2} B^\mu .
\label{improve}
\eea
This is an improvement that is consistent with the diffeomorphism, $R$-symmetry and scale Ward identities since these improvement terms obey
\bea
\nabla^{(0)}_\mu r^\mu &=& 0 \cr
g^{(0)\mu\nu} \tau_{\mu\nu} &=& 0 \cr
\nabla^{(0)\mu} \tau_{\mu\nu} &=& F^{(0)}_{\mu\nu} r^\nu .
\eea
In addition, the completely traceless and divergenceless improvement for the energy-momentum tensor can be obtained from the addition of the finite boundary action term
\bea
\Gamma_{\alpha} &=& \frac{\alpha}{32} \int_{\rho = \epsilon} d^4 x \sqrt{-\gamma}\left[ R_{\mu\nu}(\gamma) R^{\mu\nu}(\gamma) -\frac{1}{3}R^{2}(\gamma) \right] \cr
 &=&\frac{\alpha}{32} \int_{\rho = 0} d^4 x \sqrt{-g^{(0)}}\left[ R_{\mu\nu}^{(0)} R^{(0)\mu\nu} -\frac{1}{3}R^{(0)2}\right] ,
\label{ImprovementCounter-term3}
\eea
with $\alpha$ an arbitrary constant.  The new improvement term for the energy-momentum tensor becomes
\be
\tau_{\mu\nu}^{(\alpha)} = \frac{2}{\sqrt{-g^{(0)}}}\frac{\delta \Gamma_{\alpha}}{\delta g^{(0)\mu\nu}}= \alpha \left[ h_{\mu\nu} +\frac{1}{4} \kappa T^{(2)}_{\mu\nu}\right] .
\label{alphaimproved}
\ee
It also is consistent with the current Ward identities as $r^{(\alpha)}_\mu =0$ and
\bea
\nabla^{(0)\mu} \tau_{\mu\nu}^{(\alpha)} &=& 0 \cr
g^{(0)\mu\nu} \tau_{\mu\nu}^{(\alpha)} &=& 0.
\eea
Hence the final expressions for the improved energy-momentum tensor and $R$-symmetry current is obtained from equations (\ref{finalcurrents}), (\ref{improve}) and (\ref{alphaimproved})
\bea
r^\mu_{\rm Improved} &=& R^\mu + r^\mu = -2R_5 Z_A \left( A^{(2)\mu} +\left(1+\ln{\tau}\right) B^\mu \right) -4C E^{\mu\nu\rho\sigma} (g^{(0)}) A^{(0)}_\nu F^{(0)}_{\rho\sigma} \cr
\tau^{\rm Improved}_{\mu\nu} &=& \theta_{\mu\nu} + \tau_{\mu\nu} + \tau_{\mu\nu}^{(\alpha)} \cr
 &=&\frac{R_5^3}{\kappa} \left[ 2 g^{(4)}_{\mu\nu} + \left(1+2\ln{\tau} +\alpha  \right)h_{\mu\nu}- g^{(2)}_{\mu\nu} g^{(0)\rho\sigma} g^{(2)}_{\rho\sigma} -g^{(0)}_{\mu\nu} \left( 2 g^{(0)\rho\sigma} g^{(4)}_{\rho\sigma} + g^{(2)\rho\sigma} g^{(2)}_{\rho\sigma}\right) \right]\cr
 & &-\frac{R_5^3}{4\kappa} \left[ R^{(0)}_{\mu\lambda} R^{(0)\lambda}_{~~~\nu} +R^{(0)}_{\lambda\mu\rho\nu} R^{(0)\lambda\rho} -\frac{1}{2} \nabla^{(0)2} R^{(0)}_{\mu\nu} +\frac{1}{6}\nabla^{(0)}_\mu  \nabla^{(0)}_\nu R^{(0)} +\frac{1}{12} g^{(0)}_{\mu\nu}\nabla^{(0)2} R^{(0)2}  \right.  \cr
 & &\left. +\frac{1}{2} R^{(0)}_{\rho\sigma} R^{(0)\rho\sigma} g^{(0)}_{\mu\nu}  -\frac{1}{2} R^{(0)}R^{(0)}_{\mu\nu}   \right] \cr
 & & -4D E^{\xi\rho\sigma\lambda} (g^{(0)}) \left\{ \nabla^{(0)}_\alpha \left[ A_\lambda^{(0)} \left( g^{(0)}_{\xi\mu} R^{(0)\alpha}_{~~~\nu\rho\sigma} + g^{(0)}_{\xi\nu} R^{(0)\alpha}_{~~~\mu\rho\sigma}\right) \right] \right.  \cr
 & &\left. +2\nabla^{(0)}_\rho \left[A^{(0)}_\lambda \left(g^{(0)}_{\xi\mu} g^{(2)}_{\nu\sigma} + g^{(0)}_{\xi\nu} g^{(2)}_{\mu\sigma} \right) \right] 
 -2 A^{(0)}_\lambda \left(g^{(0)}_{\xi\mu}\nabla^{(0)}_\rho g^{(2)}_{\nu\sigma} + g^{(0)}_{\xi\nu} \nabla^{(0)}_\rho g^{(2)}_{\mu\sigma} \right)\right\} \cr
 & &+\frac{\alpha}{4} \kappa \frac{1}{R^2_5}\left[ Z_A F^{(0)}_{\mu\rho} F^{(0)\rho}_{~~~~~\nu} +\frac{
Z_A}{4} g^{(0)}_{\mu\nu} F^{(0)}_{\rho\sigma} F^{(0)\rho\sigma} \right]  .
\eea

\section{Conclusions \label{section4}}

The holographic $R$-current and $R$-charges require additional normalizations in order for the R-current to belong to the same ${\cal N} =1$ SUSY multiplet as the holographic energy-momentum tensor.  In addition, a finite wavefunction renormalization of the $U(1)_R$ gauge field (c.f. Eq. (\ref{wavefunction})) was needed in order that the holographic contribution of the field strength to the trace anomaly equation (\ref{SCFTtraceanomaly}) is consistent with the boundary SCFT trace anomaly.  The general structure of the trace anomaly including the subleading corrections was reviewed in Appendix A for the generic non-SUSY case.  The $U(1)$ field strength subleading contribution was obtained for generic 4 derivative terms in the action which was used in the introduction to fix the gauge field normalization in the supersymmetric case.  

The subleading mixed gravitational field-$U(1)_R$ gauge field Chern-Simons term was added to the action as it gave rise to subleading gravitational contributions to the $R$-anomaly.  The modifications to the near boundary solutions to the field equations were then obtained along with the boundary counter-terms and normalization required by holographic renormalization.  Once the on-shell action was obtained, the Brown-York energy-momentum tensor and $R$-symmetry current were constructed.  The near boundary solutions were used to secure the Ward identity obeyed by the currents as
\be
\nabla^\mu \theta_{\mu\nu} = F^{(0)}_{\mu\nu} R^\mu + A^{(0)}_\nu \nabla_\mu R^\mu .
\ee
along with the trace and $R$- anomalies
\bea
\theta^\mu_{~~\mu} &=& \frac{R_5^3}{\kappa}\left[ \frac{1}{8}\left( R^{(0)}_{\mu\nu} R^{(0)\mu\nu} -\frac{1}{3}R^{(0)2} \right) +\frac{\kappa Z_A}{4 R_5^2} F^{(0)}_{\mu\nu} F^{(0)\mu\nu} \right] \cr
\nabla_\mu R^\mu &=& C\, E^{(0)\mu\nu\rho\sigma} F^{(0)}_{\mu\nu} F^{(0)}_{\rho\sigma} + D\, E^{(0)\mu\nu\rho\sigma} R^{(0)\xi\zeta}_{\mu\nu} R^{(0)}_{\xi\zeta\rho\sigma}  .
\eea
The Ward identities for diffeomorphism invariance of the action then followed as given in equations (\ref{diffeo}) and (\ref{diffeo2}).  Likewise the $R$-symmetry transformation of the action was obtained in equations (\ref{Rvar}) and (\ref{Rvar2}) while the Weyl scaling of the action followed in equations (\ref{Weyl1}) and (\ref{Weyl2}).  Since the Ward identities provide the interpretation of the holographic currents as the energy-momentum tensor and $R$-symmetry current, improvements to the currents were constructed which left the Ward identities unchanged.  As explicitly demonstrated, the improvements were expressed as additional finite boundary terms in the action.

Lastly, the fermionic gravitino sector of the conformal supergravity action can be included \cite{Balasubramanian:2000pq}, although left here for future work.  The boundary ${\cal N}=1$ SCFT includes the supersymmetry current in the supercurrent multiplet and the superconformal anomaly, given by the $\gamma$ trace of the supersymmetry current, as part of the anomaly multiplet.  The central charges $a$ and $c$ also describe the superconformal anomaly.  The holographic supersymmetry currents can be constructed and their divergence and trace determined.  The Ward identities will then include these fermionic currents as well, while anomaly matching will provide additional consistency checks for the $AdS$/gauge duality.

\begin{acknowledgments}
The authors thank Martin Kruczenski and Georgios Michalogiorgakis for helpful and enlightening discussions.  The work of TEC and STL was supported in part by the U.S. Department of Energy under grant DE-FG02-91ER40681 (Theory).  The work of TtV was supported in part by a Cottrell Award from the Research Corporation and by the NSF under grant PHY-0758073.  
\end{acknowledgments}
~\\

\appendix
\section{Subleading Holographic Anomalies \label{appendixA}} 

The holographic calculation of the  $U(1)$ and trace anomalies for a generic $U(1)$ gauge field coupled to a gravitational theory with a cosmological constant in five dimensions, including four derivative terms, is considered. The bosonic part of the gauged supergravity action relevant for the holographic trace anomaly calculation is obtained by imposing the appropriate SUSY relationships among the generic parameters as presented below. The leading, two derivative part of the action contributing to the trace anomaly takes the form
\begin{eqnarray}
\Gamma_{\rm Leading}^{(1)} & = & \int d^5x \,{\rm det} E \left( \frac{1}{2 \kappa} R + \Lambda - \frac{Z_A}{4} F_{MN} F^{MN} \right).
\end{eqnarray}
The relation between the gravitational constant, the cosmological constant, and the radius of $AdS$ space follows from the Einstein equation as
\begin{eqnarray}
\kappa \Lambda R_5^2 = -6.
\end{eqnarray}
Three types of four derivative terms are relevant for the holographic calculation of the trace anomaly. The first type includes the curvature-squared terms \cite{Nojiri:1999mh}-\cite{Cremonini:2008tw}
\begin{eqnarray}
\Gamma_{\rm Subleading}^{(1)} & = & \int d^5x \,{\rm det} E \left[ \alpha \, R^2 + \beta \, R_{MN} R^{MN} + \gamma \, R^{MNKL} R_{MNKL} \right].
\end{eqnarray}
Including these terms, the modified Einstein equation still allows Anti-de Sitter space as a solution, but results in an altered relationship between the gravitational constant, the cosmological constant, and the radius of $AdS$ space, now involving also the parameters $\alpha$, $\beta$, and $\gamma$, as
\begin{eqnarray}
6 \frac{R_5^3}{2 \kappa} + \frac{1}{2} \Lambda R_5^5 + 40 \alpha R_5 + 8 \beta R_5 + 4 \gamma R_5 & = & 0.
\end{eqnarray}
Note that for the special values of the parameters $\alpha=1/6$, $\beta=-4/3$, and $\gamma=1$, the three curvature-squared terms combine to yield the square of the Weyl tensor, and the relationship between the gravitational constant, the cosmological constant, and the radius of AdS space is seen to reduce to the one that is obtained in the absence of the curvature-squared terms. A second type of four derivative terms couples the square of the gauge field strength tensor to the curvature tensors,
\begin{eqnarray}
\Gamma_{\rm Subleading}^{(2)} & = & \int d^5x \,{\rm det} E \left[ \delta \, R F_{MN} F^{MN} + \xi \, R^{MN} F_{MK} F_{NL} g^{KL} + \zeta \, W^{MNKL} F_{MN} F_{KL} \right],\cr
 & & 
\end{eqnarray}
and a third type includes terms that involve the square of the gauge field strength tensor and two additional derivatives,
\begin{eqnarray}
\Gamma_{\rm Subleading}^{(3)} & = & \int d^5x \,{\rm det} E \left [\mu \, F_{MN} \nabla^N \nabla_R F^{MR} + \nu \, \nabla^M F^{NR} \nabla_M F_{NR} + \sigma \, \nabla^M F^{NR} \nabla_N F_{RM} \right].\cr
 & & 
\end{eqnarray}
The holographic trace anomaly is determined by the logarithmically divergent part of the on-shell action as
\begin{eqnarray}
\Gamma_{\rm on-shell} & = & \dots + \frac{1}{2} \ln\left( \frac{\epsilon}{R_5^2}\right)  \int dx^4 \sqrt{- {\rm det} g^{(0)}} {\cal A} + \dots  ,
\end{eqnarray}
with the regulating near boundary surface located at $\rho =\epsilon$ and with minimal subtraction at $\rho = R_5^2$.  For generic values of the parameters, the holographic  trace anomaly thus obtained reads (see \cite{Nojiri:1999mh}-\cite{Cremonini:2008tw} for the gravitational contribution)
\begin{eqnarray}
{\Theta^\mu}_\mu & = & \left( \frac{1}{8}  \frac{R_5^3}{2 \kappa} + 5 \alpha R_5 + \beta R_5 - \frac{1}{2} \gamma R_5 \right) W^2 +  \nonumber \\
& &  \left(  - \frac{1}{8}\frac{R_5^3}{2 \kappa} - 5 \alpha R_5 -\beta R_5 - \frac{1}{2} \gamma R_5 \right) {R_{\rm GB}}^2 + \nonumber \\
& & \left(\frac{1}{4} R_5 Z_A -20 \frac{\delta}{R_5} - 4 \frac{\xi}{R_5}  + 6 \frac{\nu}{R_5} - 3 \frac{\sigma}{R_5} \right) F_{\mu\nu}^{(0)} F^{(0)\mu\nu},
\end{eqnarray}
where the square of the Weyl tensor and the Einstein-Gauss-Bonnet invariant are defined as
\begin{eqnarray}
W^2 & = &  W_{\mu\nu\kappa\lambda}^{(0)} W^{(0)\mu\nu\kappa\lambda} = R_{\mu\nu\kappa\lambda}^{(0)} R^{(0)\mu\nu\kappa\lambda} - 2 R_{\mu\nu}^{(0)} R^{(0)\mu\nu} + \frac{1}{3} R^{(0)2},\nonumber \\
{R_{\rm GB}}^2 & = & R_{\mu\nu\kappa\lambda}^{(0)} R^{(0)\mu\nu\kappa\lambda} -4 R_{\mu\nu}^{(0)} R^{(0)\mu\nu}  + R^{(0)2}.
\end{eqnarray}
The anomaly in the divergence of the $U(1)$ current $J^\mu$ is holographically obtained from the  variation  of the action under the $U(1)$ gauge transformation $\delta A_T = \partial_T  \Omega(x)$. The relevant leading, two derivative pure Chern-Simons term in the action is
\begin{eqnarray}
\Gamma_{\rm Leading}^{(2)} & = & C\, \int d^5x \,{\rm det} E\,  E^{MNRST} F_{MN} F_{RS}\, A_T,
\end{eqnarray}
while the subleading, four derivative mixed Chern-Simons term takes the form
\begin{eqnarray}
\Gamma_{\rm Subleading}^{(4)} &  =  & -D \int d^5 x \, {\rm \det} E\, E^{MNRST} R^X_{~~YMN} R^Y_{~~XRS} \, A_T .
\end{eqnarray}
The $U(1)$ anomaly is obtained from
\begin{eqnarray}
\delta \Gamma & = & - \int d^4 x\, \Omega (x)\, {\cal A}_R ,
\end{eqnarray}
resulting in
\begin{eqnarray}
\sqrt{-g}\nabla_\mu J^\mu & = & C \,  \epsilon^{\mu\nu\rho\sigma} F^{(0)}_{\mu\nu} F^{(0)}_{\rho\sigma} +D\,  \epsilon^{\mu\nu\rho\sigma} R^{(0)\xi\zeta}_{~~~~~\mu\nu} R^{(0)}_{\xi\zeta\rho\sigma}  .
\end{eqnarray}
All of the above terms appear in the bosonic sector of the five-dimensional ${\cal N} = 2$ conformal supergravity action \cite{Gunaydin:1984ak} further extended with four derivative terms \cite{Hanaki:2006pj,Cremonini:2008tw}, $\Gamma=\Gamma^{(1)}_{\rm Leading} +\Gamma^{(2)}_{\rm Leading}+\Gamma^{(1)}_{\rm Subleading}+\cdots +\Gamma^{(4)}_{\rm Subleading}$, with now the gauge field $A_\mu$ corresponding to the gravi-photon and the $U(1)$ current $J^\mu$ equal to the R-current. In terms of the gravitational constant $2\kappa$, the cosmological constant $\Lambda$,  the wavefunction renormalization factor $Z_A$, and the parameter $\gamma$ that sets the scale for the subleading terms, supersymmetry forces the remaining parameters to take the values
\begin{eqnarray}
C & = &  \frac{1}{12 \sqrt{3}} Z_A^{\frac{3}{2}} \sqrt{2 \kappa} \left( 1 - \frac{8}{3} \gamma \Lambda (2 \kappa)^2 \right) \nonumber \\
D & = & - \frac{1}{2\sqrt{3}}  \gamma  Z_A^{\frac{1}{2}} \sqrt{2 \kappa} \nonumber \\
\alpha & = & \frac{1}{6} \gamma \nonumber \\
\beta & = & -\frac{4}{3} \gamma \nonumber \\
\delta  & = & \frac{1}{6}  \gamma Z_A (2\kappa) \nonumber \\
\xi  & = & - \frac{8}{3}  \gamma Z_A (2 \kappa) \nonumber \\
\mu & = & - 4 \gamma Z_A (2 \kappa) \nonumber \\
\nu & = & - 2 \gamma Z_A (2 \kappa) \nonumber \\
\sigma & = &-  2 \gamma Z_A (2 \kappa).
\end{eqnarray}

\newpage
\end{document}